\documentclass[twocolumn,pra,superscriptaddress]{revtex4-1}

\usepackage{amsmath,mathtools}
\usepackage[table]{xcolor}
\usepackage{subfigure}
\usepackage{hyperref}
\usepackage{todonotes}
\usepackage{physics}
\usepackage[normalem]{ulem}
\usepackage{physics}
\newcommand{\mi}{\mathrm{i}}
\newcommand{\fleq}[1]{Eq.~\eqref{#1}}

\begin{document}
\title{Distillation of optical Fock-states using atom-cavity systems.}

\author{G.~P.~Teja}
\email[]{teja4477@gmail.com}
\affiliation{Nanyang Technological University 21$\text{,}$ Nanyang Link 637371$\text{,}$ Singapore}

\author{Chanchal}
\email[]{chanchal4592@gmail.com}
\affiliation{Department of Physical Sciences$\text{,}$ Indian Institute of Science Education and Research$\text{,}$ Mohali$\text{,}$ Punjab 140306$\text{,}$ India }


\begin{abstract}
Fock states are quantized states of electromagnetic waves with diverse applications in quantum optics and quantum communication. However, generation of arbitrary optical Fock states still remains elusive. Majority of Fock state generation proposals rely on precisely controlling the atom-cavity interactions and are experimentally challenging.  
We propose a scheme to  distill an optical Fock state from a coherent state. A conditional phase flip (CPF) with arbitrary phase is implemented  between the atom and light.
The CPF along with the unitary rotations and measurements on the atoms enables us to distil required Fock-state. As an example, we show the distillation of Fock-sate $\ket{100}$.
\end{abstract}


\maketitle

\section{introduction}
The second quantization of electromagnetic fields reveals the equivalence of light modes with harmonic oscillators \cite{loudon2003}. 
The eigenstates of these oscillators are called the Fock states and they 
describe the number of photons present in a specific mode of a light field.  Due to their highly non-classical nature, they are extremely useful in various areas such as quantum metrology, quantum key distribution (QKD) protocols, and quantum computation \cite{2007_kok,ralph_2006}. However, their quantum nature also makes it difficult to produce them. 
In existing experiments, optical Fock states are produced using parametric down-conversion and photon number detectors. These techniques have been shown to produce Fock states up to 5 photons \cite{2006_waks,2019_Tiedau,cooper2013,branczyk2010,Lingenfelter_2021}, however, it is still a challenging task to produce higher Fock states in the optical regime.


Atom-cavity systems have been widely studied for deterministic generation of optical Fock states, 
Fock states  are created inside the cavity by controlling the interaction between atoms and cavities \cite{2003_brown,xia2012,1996_law}. However, this requires precise control which is limited by the coherence time of the system. As the number of photons increases, control becomes harder and the fidelity of generated Fock states decreases \cite{2020uria}, making it difficult to generate high photon number states.

In addition to controlling the dynamics of atom-cavity systems, schemes based on feedback mechanisms have also been proposed for the deterministic generation of Fock states \cite{2009_dots,2006_geremia}
. These schemes involve using a controller to probe the cavity mode with weak measurements, which provide information about the cavity field. Based on the information obtained from these measurements, an actuator applies feedback to the cavity. By repeatedly measuring and applying feedback, the cavity state can be prepared in any desired state. This method has been experimentally verified using Rydberg atoms and superconducting cavities \cite{2011_sayrin}.

Further, another method for Fock-state sorting has been demonstrated using chiral coupling of a two-level system to a waveguide. In this method, the light acquires a Fock-number-dependent delay, and the incident pulse is sorted temporally by its Fock numbers \cite{2020_Mahm,molmer2022}.

In this article, we propose a protocol to distil Fock states from a coherent state. It is based on repeated reflection of light from the atom-cavity system. Unlike the existing protocols, this doesn't require precise control or feedback mechanisms. Also the Fock states in this scheme are generated outside the cavity avoiding further extraction which can effect the statistics of the light.

In Sec.~\ref{sec:atm-ph} we discuss the  conditional phase flip (CPF) between the atom and the light mode and study effects of cavity-light detuning on the CPF. 
Then in Sec.~\ref{sec:dist} the atom-photon phase gate along with unitary rotations and measurements are used to distil a Fock-state from a coherent state. Finally we conclude by discussing possible implementations.

\section{ Atom-photon gate} \label{sec:atm-ph}
\begin{figure}[!tbh]
\includegraphics[width=8cm,height=3.5cm]{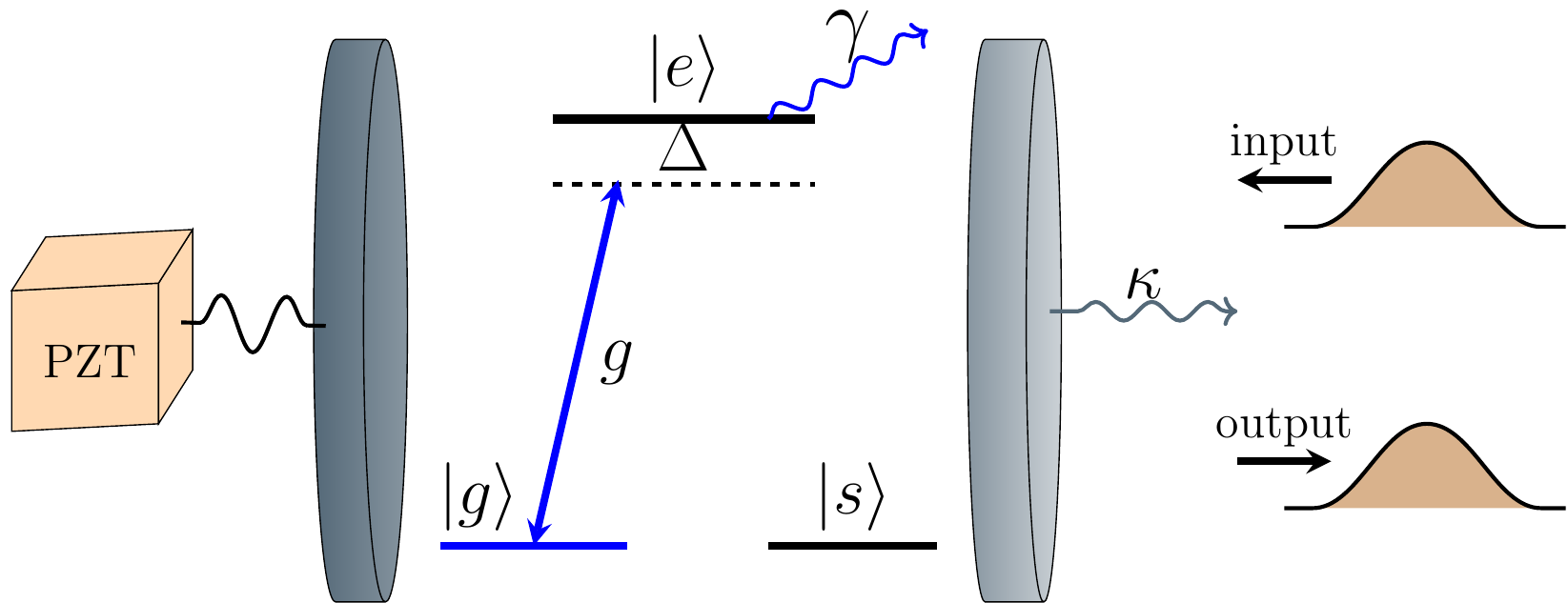}
\caption{Atom cavity system for the CPF.
A $\Lambda$-type atom with only $\ket{e}\leftrightarrow\ket{g}$ transition interacts with the cavity mode. $g$ and $\kappa$ are the coupling strength and the cavity decay rate of the cavity mode. $\Delta$ is detuning of the cavity mode with input light and the atom. The second mirror is attached to a Piezo actuator to tune the cavity resonant frequency.}
\label{fig:cav}
\end{figure}

Atom-photon gates aim to create a CPF between an atom and a flying photon. 
Typically this is confined only to a phase of $\pi$ as this is sufficient to implement a general unitary operation on atom-photon and photon-photon systems.
To obtain a conditional phase shift, we consider a three level atom with ground states $\ket{g}$ and $\ket{s}$, and an excited state $\ket{e}$ as shown in with two dipole allowed transitions ($\ket{e}\leftrightarrow\ket{g}$ and $\ket{e}\leftrightarrow\ket{s}$), see Fig.~\ref{fig:cav}. Here, only one of them $(\ket{e}\leftrightarrow\ket{g})$ is coupled to the cavity mode. The Hamiltonian of such system can be written as
\begin{align}\label{eq:ham}
H = \hbar \omega_c a^\dagger a +\hbar\omega_{eg}\dyad{e}+\hbar g(\sigma_{eg}\hat{a}
+\sigma_{ge}\hat{a}^\dagger),
\end{align}
$\omega_{eg}~(\omega_c)$ is the transition frequency for the atom (cavity), $g$ is the atom cavity  coupling strength. $\sigma_{eg} = \dyad{e}{g}$ is the atomic operator and $\hat{a}$ is the cavity mode operator.
Although we are considering a single two-level atom interacting with the cavity mode, following results can be extend to dark states in $N$ two-level systems and $N$ atoms under Rydberg blockade conditions \cite{Sun_2018,hao2019}.

On solving the \fleq{eq:ham} for the dynamics yields the following reflection coefficients $r_{(1,0)} =\hat{a}_{\text{in}}/ \hat{a}_{\text{out}}$~(see Appendix.~\ref{app:coff} for details)
\begin{align}
r_1=1-\dfrac{2}{(1+4C)-\mi \Delta}, && 
r_0=1-\dfrac{2}{1-\mi \Delta}, \label{eq:rf}
\end{align}
where $\hat{a}_{\text{in}}$ ($\hat{a}_{\text{out}}$) is the operator for the input (output) mode.
$r_1 $ is the reflection coefficient for the coupled transition. 
$r_0$ is the reflection coefficient for the decoupled transition (empty cavity) \cite{1984_gadcol}. 
$C=g^2/\kappa \gamma$ is the cooperativity.
$\Delta= 2\Delta_c/\kappa$,  $\Delta_c= \omega_{c}-\omega_L$ with $\omega_L$ being the mean frequency of the input pulse.
Now by assuming $C \gg 1$  gives $r_1 \simeq 1$ and $r_0$ can be any phase $e^{\mi \phi}$. For example, numerically solving $r_0$ for $\phi= \qty{\pi/2,\pi/4,\pi/8,\pi/16}$ gives $\Delta= 
\qty{-1,-2.4142,-5.0273,-10.1531} $.

On using \fleq{eq:rf} under strong atom-cavity coupling, the operation for the atom-cavity reflection can be written as~(see Appendix.~\ref{app:coff})
\begin{align}\label{eq:dhat}
\hat{\mathcal{D}}(\phi) = \dyad{g} \otimes \mathcal{I} + \dyad{s} \otimes \exp{\mi \phi  \hat{\Large n}}.
\end{align}
Note that cavity-light resonance  $\Delta=0$ gives $r_0 =-1 $ ($\hat{\mathcal{D}}(\pi)$). This resonance condition has been exploited to generate atom-photon gates between atom and a single photon \cite{2015_reiserer}. Further \fleq{eq:dhat} transforms a coherent state into
\begin{align}\label{eq:apo2}
\hat{\mathcal{D}}(\pi) [\psi_a  \otimes \ket{\alpha}] 
= \dfrac{1}{\sqrt{2}} (\ket{g} \ket{\alpha}+\ket{s}\ket{-\alpha} ) ,
\end{align}
where $\psi_a = \frac{1}{\sqrt{2}} (\ket{g}+\ket{s} )$.
Performing a $\small{\pi/2}$-rotation and a measurement on the atom,  projects the light to a cat-state. This has been experimentally verified by trapping
$\prescript{87}{}{\text{Rb}}$ to generate cat states with strength $\alpha = 1.4$ \cite{hacker2019}. 

\begin{figure}
\includegraphics[width=8.5cm,height=4.5cm]{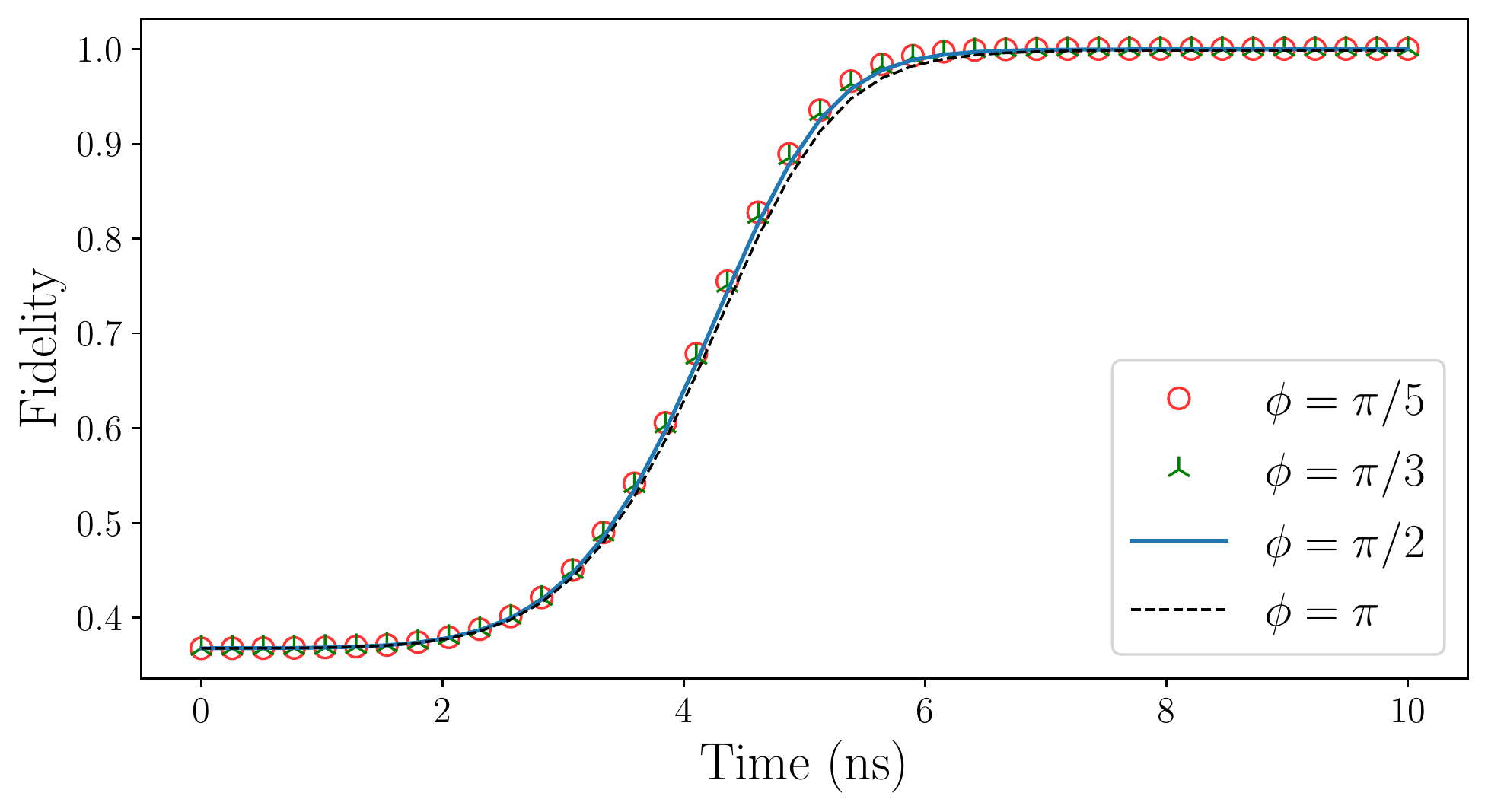}
\caption{$\hat{\mathcal{D}}(\phi)$ operation attained from the steady state dynamics of atom-cavity system. We set $\qty(g, \kappa, \gamma) = 2\pi\times\qty(16,~5,~0.05)~\text{GHz}$ for the numerical simulations. }\label{fig:fid}
\end{figure}

The reflection coefficients are derived under the assumption that the atom-cavity system quickly attains a steady state. Though this is true on resonance \cite{2020_kii}, detuning can effect the dynamics. 
In order to verify that steady states are attained even with detuning, we use input-output theory with quantum pulses to solve for the complete dynamics \cite{2019_inpout,2020_kii}. Upon reflection of light from the atom-cavity system, the following transformation is expected:
\begin{align}\label{eq:exps}
\psi_a \ket{\psi}_\text{in} \ket{0}_\text{out} \xrightarrow[\text{tion}]{\text{reflec}} 
\ket{0}_\text{in} \dfrac{\ket{g} \ket{\psi}_\text{out} +\ket{s} e^{\mi\phi \Huge{\hat{n}}}\ket{\psi}_\text{out}}  {\sqrt{2}}
\end{align}
and  $\ket{\psi}_\text{in,out} = \sum_n C_n \ket{n}$ represents the quantum state of the input and output light modes.

In Fig.~\ref{fig:fid} we plot the fidelity for various phases. Here an input light is reflected from the atom-cavity system and the quantum state of the atom and the output light is numerically obtained.  Then the fidelity is evaluated  between the obtained and the expected output state \fleq{eq:exps} (see Appendix.~\ref{app:verif} for more details). From the plot it is clear that  reflection from atom-cavity yields the operation $\hat{\mathcal{D}}(\phi)$ on the output light mode and  
a steady is obtained even in the presence of cavity-light detuning. 

\section{distillation of Fock states} \label{sec:dist}
We now use the phase shift operation $\hat{\mathcal{D}}(\phi)$, local unitary and measurement  on atom  to distil a Fock state. The  protocol consists of the following steps:
\begin{enumerate}
\item To prepare the Fock state $\ket{\mathcal{A}}$, we start with a coherent state of amplitude $\alpha = \sqrt{\mathcal{A}}$.
\item  Atom is prepared in the state $\psi_a = \frac{1}{\sqrt{2}}(\ket{g}+\ket{s})$  
\item  The light is reflected from the atom-cavity setup to perform $\hat{\mathcal{D}}(\phi)$.
\item  A unitary rotation $U_a$ is performed on the atom \cite{2017vitanov}.
$
U_a [\theta] = \dfrac{1}{\sqrt{2}} \mqty[e^{\mi  \theta}   &  1 \\
1 & -e^{-\mi  \theta} ] ,
\mqty{
&\ket{g} \to \frac{1}{\sqrt{2}} (e^{\mi  \theta} \ket{g}+\ket{s})\\
&\ket{s} \to \frac{1}{\sqrt{2}} ( \ket{g}-e^{-\mi  \theta}\ket{s})
}
$ 
\item  Finally a measurement is performed on the atom \cite{hacker2019}.\\
\end{enumerate}
Choosing $\alpha = \sqrt{\mathcal{A}}$ is not a stringent condition, this choice is based on the fact that coherent state has Poisson distribution with peak value at $\alpha^2$. 
The state of the ``atom+output light" after step-$3$ can be written as 
\begin{align}\label{eq:stp3}
\Psi =& \dfrac{1}{\sqrt{2}} \qty[\psi_l \ket{g} + \exp{\mi \phi \hat{n}} \psi_l \ket{s}],
\end{align}
where $\psi_l= \sum_n C_n \ket{n} $, is the state of light. Applying an appropriate $U_a[\theta]$, the atom-cavity state can always be cast in the form
\begin{align}\label{eq:stp4}
\Psi = \sum_{k} C_k \ket{k}  \ket{g} +
\sum_{l} C_l \ket{l}  \ket{s},
\end{align}
where $k \neq l$ and $ \sum_{k} \abs{C_k}^2 +  \sum_{l} \abs{C_l}^2 = 1$ and measurement on the atom gives the photonic state
\begin{align}\label{eq:stp5}
\psi_l = \sum_{k} C_k \ket{k}  ~\text{or} ~
\psi_l = \sum_{l} C_l \ket{l},
\end{align}
where $ \sum_{k} \abs{C_k}^2$ and $ \sum_{l} \abs{C_l}^2$ are the probabilities for the atomic measurements $\ket{g}$ and $\ket{s}$ .

By repeating the steps $2$ to $5$, 
it is possible to distil a general Fock state. We establish  this by  explicitly showing the distillation of  the  Fock state $\ket{100}$. 
The atom is initialized in the state $\psi_a$ and the input light in the coherent state
$\ket{\alpha=10}$. The initial state of atom and light can be written as (see Appendix.~\ref{app:sqz})
\begin{align}\label{eq:ichs}
\Psi =  \sum_{k=70}^{130}  C_k   \ket{k}  \otimes \dfrac{1}{\sqrt{2}}(\ket{g}+\ket{s}),
\end{align}
where the coherent state is approximated as $\bm{\ket{\alpha=10}}\simeq\sum_{k=70}^{130}$ $C_k   \ket{k}$. The explicit form $C_k$ is irrelevant to the distillation protocol and is only used to calculate measurement probabilities.
Performing $\hat{\mathcal{D}}(\pi)$ on $\Psi$ yields
\begin{align}
\Psi_\pi =  \sum_{k=70}^{130}  C_{k}  e^{\mi \pi \hat{n}} \ket{k} \ket{s} + \sum_{n=70}^{130}  C_k  \ket{k} \ket{g},
\end{align}
where $\Psi_l$ represents the state of output light after applying $\hat{\mathcal{D}}(l)$ operation.
Note that $\hat{\mathcal{D}}(\pi)$ is obtained on resonance i.e. $\Delta=0$.
Performing the atomic unitary $U_a[0]$ and measuring the atom in the state $\ket{g}$ removes the odd photons to give the even cat state \cite{hacker2019}
\begin{align}\label{eq:1ps}
{\Psi_\pi} =  
\sum_{k=36}^{64} C_{2k} \ket{2k},
\end{align}
here the normalization is absorbed into the $C_{2k}$ i.e. $\sum_{k=36}^{64} \abs{C_{2k}}^2 = 1$ .

Preparing the atom in $\psi_a$ and reflecting the state ${\Psi}_{\pi}$ with $\hat{\mathcal{D}}(\pi/2)$ yields 
\begin{align}
{\Psi_{\frac{\pi}{2}}} =
\sum_{k=36}^{64}  C_{2k}  \qty[(-1)^{k} \ket{s} +   \ket{g}] \ket{2k} ,
\end{align}
performing $U_a[0]$ and a measurement in $\ket{g}$ projects the photonic state to  
$
\Psi_{\pi/2} =  
\sum_{k=18}^{32}  C_{4k}    \ket{4k} 
$.
Now reflecting the state ${\Psi}_{{\pi}/{2}}$ with $\hat{\mathcal{D}}(\pi/4)$ yields
\begin{align}
{\Psi_{\frac{\pi}{4}}} = 
\sum_{k=18}^{32}  C_{4k}  \qty[(-1)^{k} \ket{s} + \ket{g}] \ket{4k},
\end{align}
applying $U_a[0]$ and a measurement in $\ket{s}$, projects the light state to
$
\Psi_{\pi/4} = 
\sum_{k=9}^{15}  C_{8k+4}  \ket{8k+4}.
$
Reflecting ${\Psi}_{\pi/4}$  with $\hat{\mathcal{D}}(\pi/8)$ yields
\begin{align}
\begin{aligned}
{\Psi_{\frac{\pi}{8}}} = 
\sum_{k=9}^{15}  C_{8k+4} \qty[(-1)^k e^{\mi \pi/2}  \ket{s} +
   \ket{g}]  \ket{8k+4} .
\end{aligned}
\end{align}
To cancel the extra factor of $\pi/2$, we perform $U_a[\pi/2]$. Now, measuring the atom in $\ket{g}$ projects the light state to
\begin{align}
\begin{aligned}
{\Psi_{\frac{\pi}{8}}} = & 
 \sum_{k=5}^{7}  C_{16k+4}  \ket{16k+4} ,\\
=&
  C_{84}  \ket{84} + C_{100}  \ket{100} + C_{116}  \ket{116}.
\end{aligned}
\end{align}
Again reflecting $\Psi_{\pi/8}$ with $\hat{\mathcal{D}}(\pi/16)$ gives 
\begin{align}
{\Psi_{\frac{\pi}{16}}} = 
\sum_{k=5}^{7} C_{16k+4} \qty[ (-1)^k e^{\frac{\mi \pi}{4} }  \ket{s} +    \ket{g} ] \ket{16k+4},
\end{align}
and finally performing $U_a[\pi/4]$ and measurement in $\ket{s}$  distils the Fock state $\ket{100}$.

\renewcommand{\arraystretch}{2}
\setlength{\tabcolsep}{0.5pt}
\begin{table}[!tbh]
\centering
\begin{tabular}{ |c|c|c|c|c|c|  }
\hline
\rowcolor{gray!30}
\multicolumn{2}{|c|}{Distilled Fock Nos.} &$\phi$ &$\theta$ &$\mathcal{M}$ &$\mathcal{P}$\\
\hline \hline
$\qty(k)_{k=70}^{130}$ & 70,\,71\dots100\dots 129,\,130  & $\pi$ & 0& $~\ket{g}~$ & \\
\rowcolor{gray!20} 
$\qty(2k) _{k=36}^{64}$ & 70,\,72\dots100\dots 126,\,128 &  $\pi/2$     & 0 &$\ket{g}$& 0.5 \\  
$\qty(4k)_{k=18}^{32}$   & 72,\,76\dots100\dots 124,\,128 & $\pi/4$      & 0 &$\ket{s}$& 0.5\\
\rowcolor{gray!20} 
$\qty(8k+4)_{k=9}^{15}$  & 76,\,84\dots100\dots 116,\,124 & $\pi/8$      & $\pi/2$ &$\ket{g}$& 0.5\\
$\qty(16k+4)_{k=5}^{7}$  & 84,~100,~116 & $~\pi/16~$  & $~\pi/4~$ &$\ket{g}$&0.5\\
\rowcolor{gray!20}
$\qty(16k+4)_{k=6}$  & 100 &  -& - &-&0.64\\
\hline
\end{tabular}
\caption{Distillation of Fock states. First column shows the sequence of distilled Fock numbers after each iteration. Second column shows the atom-cavity phase, $\hat{\mathcal{D}}(\phi)$. Other columns show the unitary operation $U_a[\theta]$ and the measurement ($\mathcal{M}$) on the atom. $\mathcal{P}$ is the probability  of the outcome after each iteration. Note that atomic is prepared in $\psi_a$ after each iteration.}\label{tab:dist}
\end{table}

The distillation protocol discussed above is  summarized in Table.~\ref{tab:dist} and it is clear that an arbitrary Fock state can be distilled by multiple iterations. 
We started with a coherent state with mean $\tiny(\expval{\hat{n}})$ and variance $\small(\expval{\hat{n}^2}-\expval{\hat{n}^2})$ of $\abs{\alpha}^2$. For  high $\mathcal{A}$, the probability distribution can be approximated with a Gaussian and  $\pm3\sqrt{\mathcal{A}}$ covers the 0.997 of the total area of a Gaussian distribution (see Appendix.~\ref{app:sqz}). Also from the Table.~\ref{tab:dist}, 
we notice that in each iteration, the total Fock numbers reduce by half. 
Thus the total number of iterations ($\mathcal{Q}$) required to distil a Fock state form the coherent state can be written as
\begin{align}\label{eq:for}
6\sqrt{\mathcal{A}}=  2^{\mathcal{Q}+1}~ \Rightarrow ~
\mathcal{Q} = \lceil{\log_2{(6\expval{\Delta\hat{n}})}}\rceil-1,
\end{align}  
where $\lceil{.}\rceil$ is the  ceiling function and $\expval{\Delta\hat{n}} = [\expval{\hat{n}^2}-\expval{\hat{n}}^2]^{1/2} $ is the standard deviation of the coherent state. Although \fleq{eq:for} is obtained assuming high coherent state amplitude, it works even for small amplitudes. Further squeezed coherent states with optimized squeezing \cite{gerry2005}
can be used to reduce the numbers of iterations to $\mathcal{Q}-1$ (see Appendix.~\ref{app:sqz}).


It is  also interesting to note that a general CPF is not required and only phase shifts
of the form $\phi = \pi/2^n$ are sufficient for the distillation protocol. Since every iteration requires an atomic measurement, the probability of success is $\sim 1/2^M$. Even though the probability of success decreases by increasing reflections, every iteration produces a highly non-classical state irrespective of the measurement outcome. Also the probability of success to generate a Fock state in the range $[\mathcal{A}-3\sqrt{\mathcal{A}},\,\mathcal{A}+3\sqrt{\mathcal{A}}]$ is unity.

Another special case of the distillation protocol is the deletion of prime numbered Fock states. Prime numbers by definition cannot be factored by any other number, this can be exploited to delete a Fock number from a given state.
For example, reflecting the coherent state $\Psi$ (\fleq{eq:ichs})
with phase $\hat{\mathcal{D}}(\pi/101)$,
performing $U_a[0]$ and measuring the atom in $\ket{g}$ gives the photonic state 
$\Psi =
\sum_{n=70}^{100}  C_n \ket{n} + \sum_{n=102}^{130} C_n \ket{n}$.


\section{Implementations and Discussion}
Atom-cavity systems are used to implement a variety of protocols \cite{2015_reiserer}.
To implement this protocol, the cavity resonant frequency can be tuned to  adjust the detuning $\Delta$. This can be achieved by attaching the cavity mirror to a peizo electric actuator \cite{mohle2013}. Besides tuning the cavity frequency, multiple atom-cavity systems with different detunings can also be considered. Waveguide QED is an effective approach to study such systems, where an array of atoms are trapped above waveguides to realize many atom-cavity systems on a single chip\cite{chang_2018}. Further, the amount of non-classicality of the output light increases after each iteration, which in-turn can make the quantum states strongly susceptible to transmission and photon losses \cite{1997_photonloss}. This discussion is beyond the scope of this article and will be studied separately.

Also, the coherent state pulses used in the distillation protocol typically have a narrow temporal widths and this can be used to realized highly non-classical bath.
When the temporal width of the pulse is much narrower than the cavity linewidth, the pulse can effectively act as a bath for an atom-cavity system. This means that the first Markov approximation is satisfied, which assumes that the system-reservoir coupling strength is frequency independent \cite{1985Gard}.

\section{Acknowledgments} We would like to thank Dr Sandeep K. Goyal and Dr. G. Krishna Teja for their helpful discussions. Chanchal acknowledges the Council of Scientific and Industrial Research (CSIR), Government of India, for financial support through a research fellowship [Award No. 09/947(0106)/2019-EMR-I].

\appendix
\section{Reflection coefficients for the atom-cavity system}\label{app:coff}
Here we derive the reflection coefficients in the main text.
The Hamiltonian of the atom cavity system is written as 
\begin{align}
H = \hbar\omega_ca^\dagger a+\hbar\omega_a\dyad{e}+\hbar g(\sigma_{eg}a+\sigma_{ge}a^\dagger), 
\end{align}
The dynamics of the system are governed by the Langevin equation and the input-output relation \citep{1984_gadcol}
\begin{align}
\begin{aligned}
\dv{\hat{\mathbf{x}}}{t} = &-\mi \comm{\hat{\mathbf{x}}}{H} 
-\Big(\comm{\hat{\mathbf{x}}}{\hat{a}^\dagger}(-\dfrac{\kappa}{2} \hat{a}+\sqrt{\kappa}\hat{a}_\text{out})
-\\
& \qquad (-\dfrac{\kappa}{2} \hat{a}^\dagger+\sqrt{\kappa}\hat{a}_\text{out}) \comm{\hat{\mathbf{x}}}{\hat{a}}\Big)\\
&-\Big(-\dfrac{\gamma}{2}\comm{\hat{\mathbf{x}}}{\sigma_{eg}} \sigma_{ge}
+\dfrac{\gamma}{2} \sigma_{eg} \comm{\hat{\mathbf{x}}}{\sigma_{ge}}\Big)
\end{aligned}
\end{align}
where  $\hat{\mathbf{x}}$ is any operator of the atom-cavity system. $\gamma$ and $\kappa$ are the decay rates of atom and cavity. 
Note that the noise term for atomic decay are omitted because $\omega_{eg}$ is in the  optical regime, hence the noise can be assumed to be vacuum. 
The dynamical equations are obtained as
\begin{align}
\dv{\hat{a}}{t}=&-\mi\omega_c\hat{a}-\mi g \sigma_{ge} -\sqrt{\kappa}\hat{a}_\text{out}+\dfrac{\kappa}{2} \hat{a},\\
\dv{\sigma_{ge}}{t}=&-\mi\omega_a\sigma_{ge}-\mi g(\sigma_{gg}-\sigma_{ee})\hat{a} +\dfrac{\gamma}{2} \sigma_{ge},
\end{align}
Transforming 
$
\mqty[\hat{a}, \sigma_{ge}, \hat{a}_\text{out}
] \to
\mqty[\hat{a}, \sigma_{ge}, \hat{a}_\text{out}
] e^{-\mi \omega_L t}
$
gives
\begin{align}
\dv{\hat{a}}{t}=&-\mi\Delta_c\hat{a}-\mi g \sigma_{ge} -\sqrt{\kappa}\hat{a}_{\text{out}}+ \dfrac{\kappa}{2} \hat{a},\\
\dv{\sigma_{ge}}{t}=&-\mi\Delta_a\sigma_{ge}-\mi g (\sigma_{gg}-\sigma_{ee}) \hat{a}+\dfrac{\gamma}{2}\sigma_{ge},
\end{align}
where $\Delta_c = \omega_c-\omega_L$ and $\Delta_a = \omega_a-\omega_L$.
The input light is on resonance with the atomic transition ($\Delta_a =0$). Also atom is assumed to be weakly excited hence $\sigma_{gg} =1$.
By setting $\pdv{\mathbf{x}}{t} = 0 $, the steady state solutions are obtained as 
\begin{align}
\hat{a} = \dfrac{-\sqrt{\kappa}}{\mi \Delta_c - \kappa/2 - \dfrac{g^2}{  \gamma/2}} \hat{a}_{\text{out}}
\end{align}
On using the input-output relation 
$
\hat{a}_{\text{out}}-\hat{a}_{\text{in}} = \sqrt{\kappa}\hat{a}
$
yields 

\begin{align}\label{aeq:rf}
r_1&=1-\dfrac{2}{1+4C-\mi \Delta}, &&
r_0=1-\dfrac{2}{1-\mi \Delta},
\end{align}
where $\Delta = 2\Delta_c/\kappa$ and the cooperativity $C={g^2}/{\kappa\gamma}$. The reflection coefficient for the non-interacting transition ($r_0$) is obtained using $C=0$.
On using \fleq{aeq:rf} an input state 
\begin{align}
\Psi_{\text{in}} =  (c_g \ket{g}+ c_s \ket{s})\otimes \sum_n c_n \dfrac{(\hat{a}_{\text{in}}^\dagger)^n}{\sqrt{n!}} \ket{0}
\end{align}
upon reflection is transformed as follows:
\begin{align}\label{aeq:out}
\Psi_\text{{out}} =  \qty (c_g \ket{g} r_1^{\hat{n}}+ c_s\ket{s} r_0^{\hat{n}}    )\otimes \sum_n c_n  \ket{n},
\end{align}
here $\abs{c_g}^2+\abs{c_s}^2 =1$ and $\abs{c_n}^2 =1$.
Form the \fleq{aeq:out} the transformation for general input state can be written as 
\begin{align}\label{aeq:dhat}
\hat{\mathcal{D}}(\phi) = \dyad{g} \otimes (r_1)^{\hat{n}} + \dyad{s} \otimes (r_0)^{\hat{n}},
\end{align}
with $r_1 \simeq 1$ and  $r_0=e^{\mi\phi}$ gives the \fleq{eq:dhat}.


\begin{table}[!tbh]
\centering
\rowcolors{2}{gray!30}{}
\begin{tabular}{ |c | c | c  |}
\hline
\rowcolor{gray!60}
$\Delta$ & $e^{\mi \phi}$ & $r_1$\\
\hline \hline
-1  & ~$e^{\mi \pi/2}$ ~      & ~$0.998+  2\times 10^{-6} \mi$  ~\\
~ -2.41421 ~ & $e^{\mi \pi/4}$ &$0.998+ 4.8 \times 10^{-6} \mi$  \\ 
-5.02734  & $e^{\mi \pi/8}$ & $0.998+ 10^{-5} \mi$  \\ 
-10.15317  & $e^{\mi \pi/16}$ &  $0.998+ 2.02 \times 10^{-5} \mi$ \\
~ -20.35547 ~& $e^{\mi \pi/32}$  &~ $0.998+  4.06 \times 10^{-5} \mi$ ~ \\
\hline
\end{tabular}
\caption{Numerical solutions for reflection coefficients, here the cooperativity $C = 250 $. $\Delta$ represents the detuning required for $r_0 = e^{\mi \phi}$. We notice that $r_1 \simeq 1$ .}
\label{tab:dist}
\end{table}

\section{Verification of general phase}\label{app:verif}

Here we discuss the verification of the general phase as described in Eq.~\eqref{eq:dhat}. For this, we use the general input-output theory with quantum pulses~\cite{2019_inpout,2020_kii}. It is based on density matrix formalism where the input and output pulses are replaced by virtual cavities coupled to the quantum system. This formalism explicitly incorporates the information of the pulse shapes and quantum states of the input and output pulses. The Hamiltonian governing the dynamics of the virtual cavities and the quantum system are given by

\begin{align}
\begin{split}
\hat{H_{\rm eff}}=\hat{H}_s+&\dfrac{\mi}{2}\sqrt{\kappa} \qty[g_u(t) \hat{a}^\dagger_u \hat{a}+g_v^*(t) \hat{a}^\dagger \hat{a}_v]\\+&\qty[g_u(t)g_v^*(t)\hat{a}_u^\dagger\hat{a}_v-h.c.], \label{molmerh}
\end{split}
\end{align}

where $H_s$  denotes the system Hamiltonian given in Eq.~\eqref{eq:ham} and $\hat{a}$ is the system cavity operator. $\hat{a}_u$ and $\hat{a}_v$ represent the input and output virtual cavity field operators with the corresponding time-dependent coupling strengths $g_u(t)$ and $g_v(t)$, respectively.

The time-dependent coupling strengths of these virtual cavities are chosen such that the input virtual cavity releases an input field with the required pulse shape $u(t)$, while the output virtual cavity acquires the output field mode with pulse shape $v(t)$.
The relation between time profiles and coupling strengths is given by
\begin{align}
g_u(t)=\dfrac{u^*(t)}{\sqrt{1-\int_0^t \abs{u'(t)}^2 dt'}}, &&
g_v(t)=-\dfrac{v^*(t)}{\sqrt{\int_0^t \abs{v'(t)}^2 dt'}}.
\end{align}

The dynamics of the system are obtained by the solving the following Lindblad master equation
\begin{align}
\dv{\rho_{usv}}{t}=\dfrac{1}{\mi \hbar}\comm{\hat{H}}{\rho_{usv}}+\mathcal{D}[\hat{L}_{\rm eff}]\rho_{us}, \label{mole1}
\end{align}
where $\rho_{usv}$ is the density matrix of the full system including the input-virtual cavity, atom-cavity system and output-virtual cavity and $\mathcal{D}[\hat{L}_{\rm eff}]$ represents the time-dependent Lindblad dissipator with
\begin{align}
\hat{L}_{\rm eff}(t)=\sqrt{\kappa}\hat{c}+g^*_u(t)\hat{a}_u+g^*_v(t)\hat{a}_v. \label{lfull}
\end{align}
The output field mode can be obtained by considering only the input virtual cavity attached with the system with the Hamiltonian given by~\cite{2019_inpout}
\begin{align}
\hat{H_{\rm 0}}=\hat{H}_s+\dfrac{\mi}{2}\qty[\sqrt{\kappa} g_u(t) \hat{a}^\dagger_u \hat{a}-\sqrt{\kappa} g^*_u(t) \hat{a}_u \hat{a}^\dagger], \label{molmerh}
\end{align}
along with the damping term given by the Lindblad operator
\begin{align}
\hat{L_0}(t)=\sqrt{\kappa}\hat{c}+g^*_u(t)\hat{a}_u\label{lu}
\end{align}
The prominent output field modes along with the amount of excitation carried by them can be obtained by calculating the eigenmode decompostion of the following two-time correlation function \cite{2020_kii}
\begin{align}
g^{(1)}(t,t')=\expval{[\hat{L}_0(t)]^\dagger\hat{L}_0(t')}\equiv \sum_i n_i v_i(t)
\end{align}
Using this, we can solve the master equation for the full system given in Eq.~\eqref{mole1} and calculate the fidelity with time for state given in Eq.~\eqref{eq:dhat}.


\onecolumngrid
\section{Distillation using  squeezed coherent states (SCS)}\label{app:sqz}
\begin{figure*}[!tbh]
\subfigure[\label{fig1}]{\includegraphics[width=5.5cm,height=4.5cm]{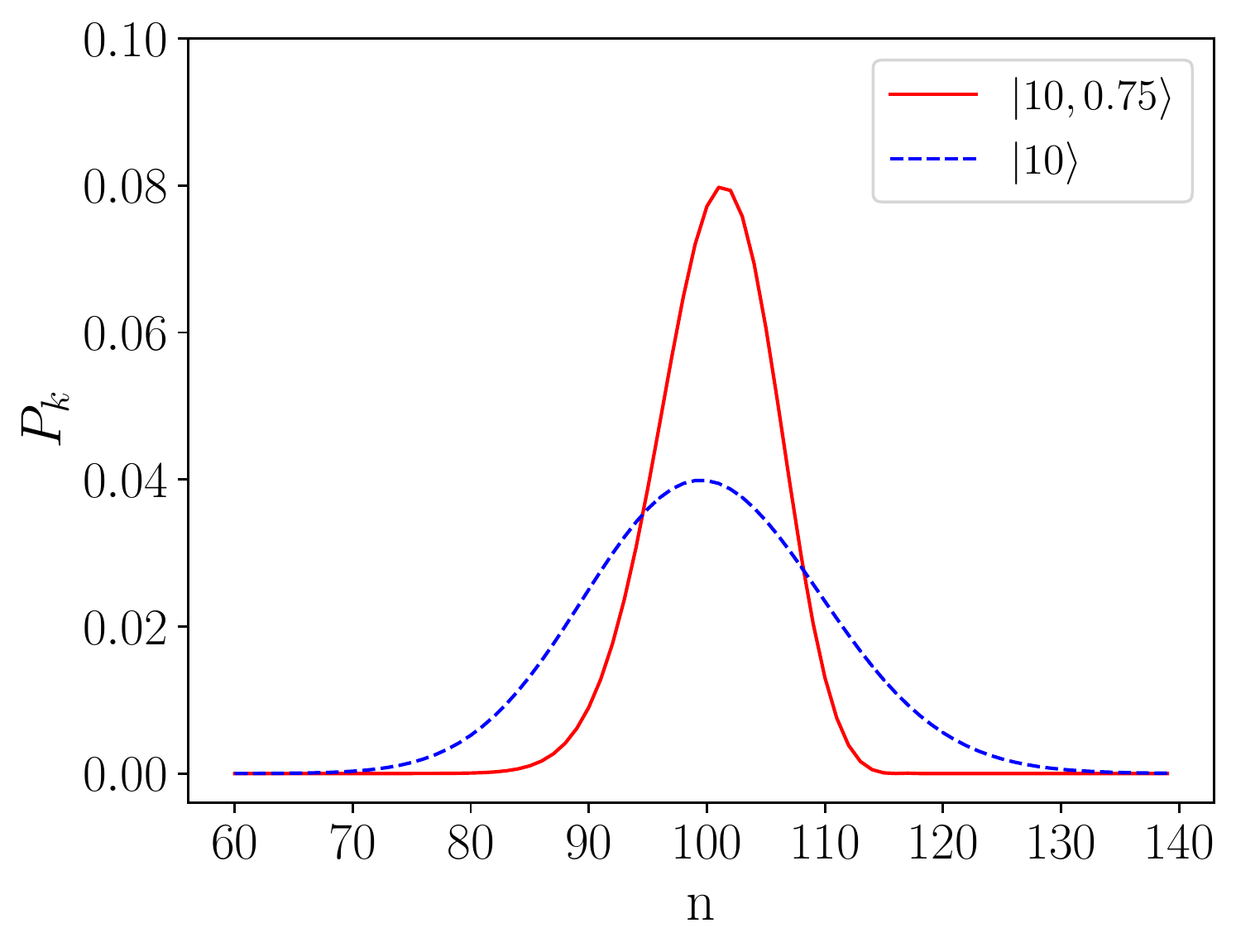}}
\subfigure[\label{fig2}]{\includegraphics[width=5.5cm,height=4.5cm]{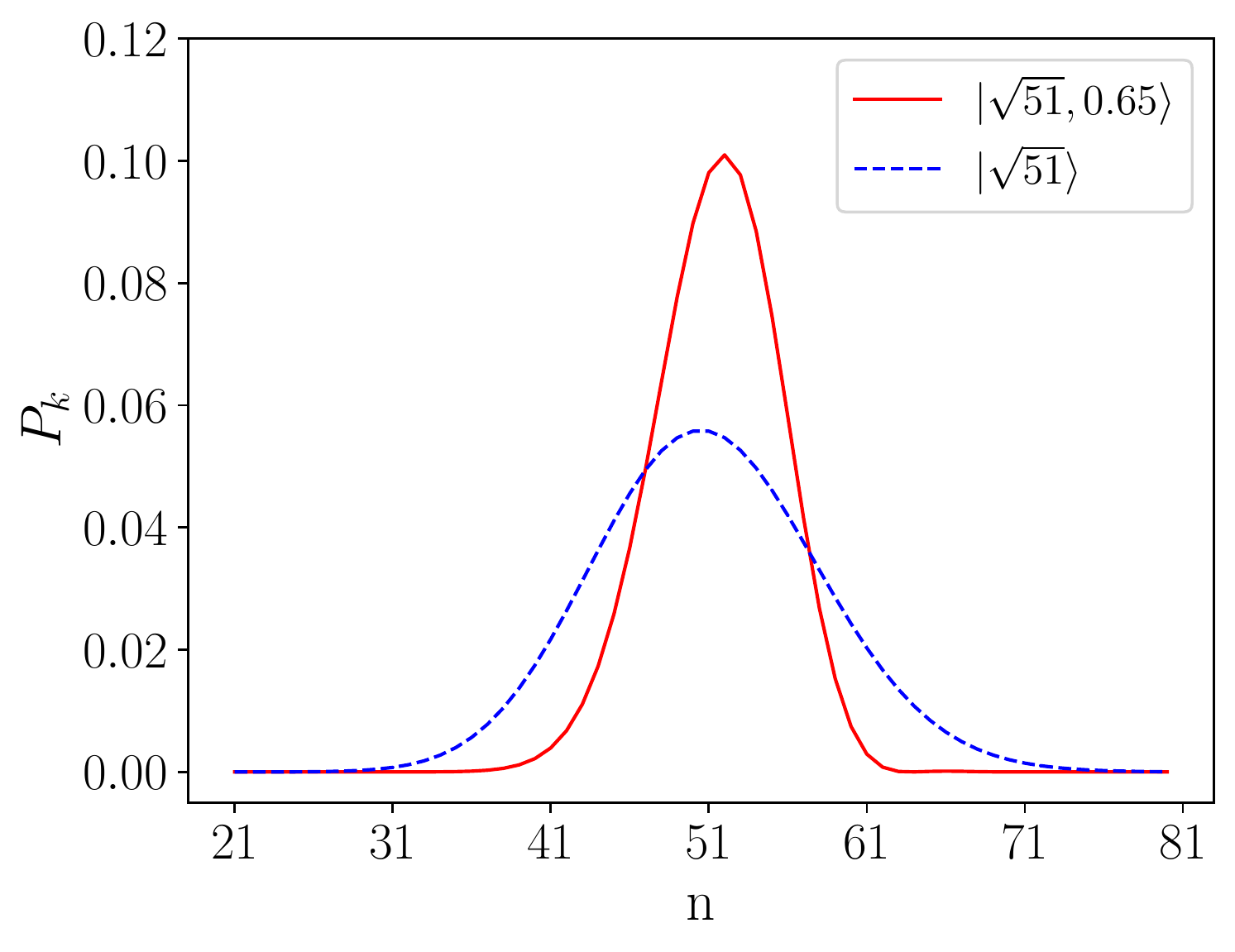}}
\subfigure[\label{fig3}]{\includegraphics[width=5.5cm,height=4.5cm]{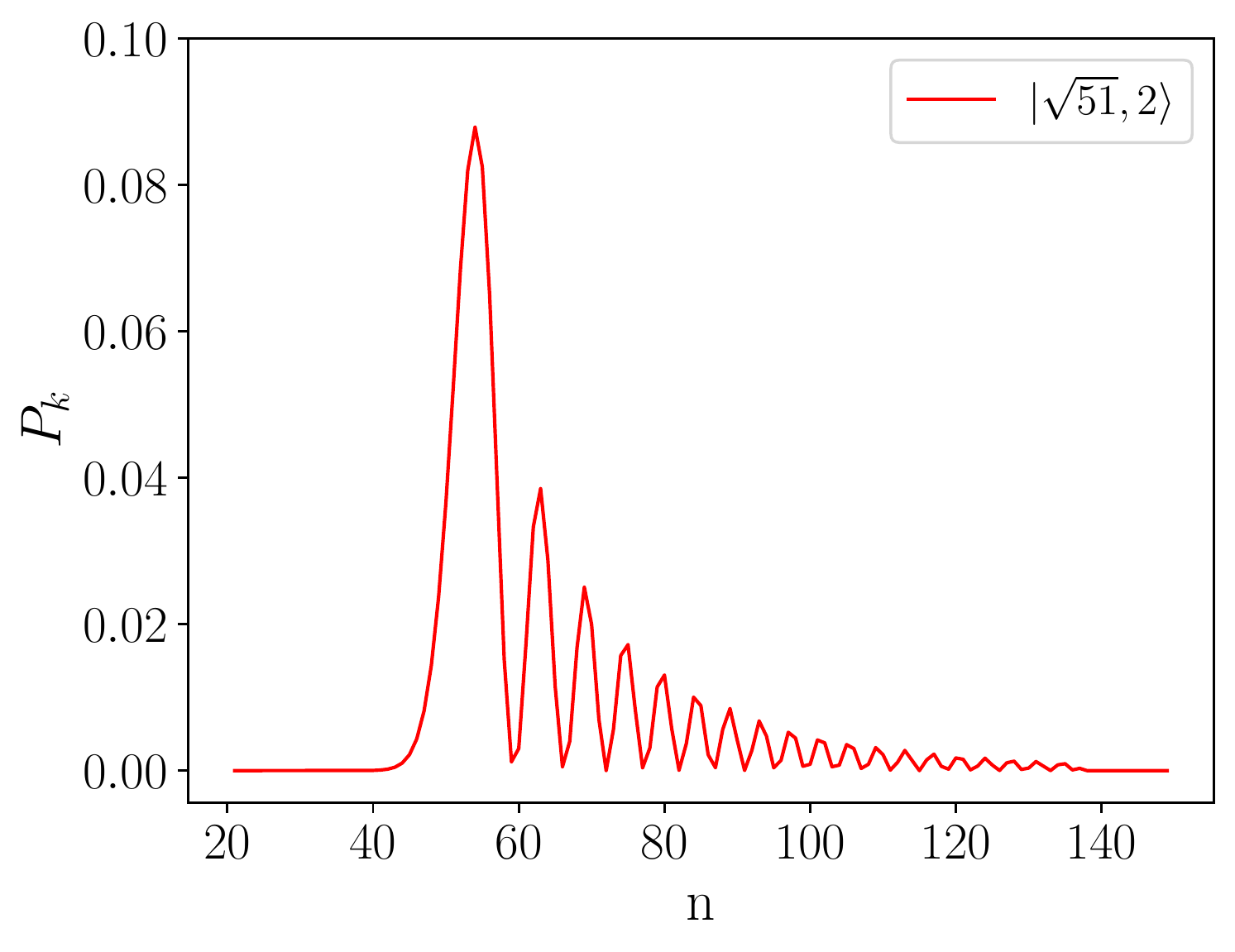}}
\caption{\subref{fig1} and \subref{fig2}: Number squeezing for the SCS. $\ket{\alpha,r}$ and $\ket{\alpha}$  represents the SCS and the coherent state. An optimal squeezing can result in quantum states which are  similar to a coherent state but with reduced variance. Due to the reduced variance SCS can be useful in minimizing iterations. \subref{fig3} The effect of photon number oscillations for higher squeezing.
}\label{fig:sts}
\end{figure*}
\begin{table*}[!tbh]
\centering
\begin{tabular}{ ||@{\quad}c@{\quad}|@{\quad} c@{\quad} | @{\quad}c@{\quad} |@{\quad} c @{\quad}|| }
\hline
Distilled Fock No. & $\phi$ & $\theta$ & $\mathcal{M}$\\
\hline \hline
36,~37\dots 51 \dots 65,~66& $\pi$ & 0& $\ket{s}$   \\ 
37,~39\dots 51 \dots 63,~65&  $\pi/2$  & $\pi/2$ &$\ket{s}$  \\  
39,~43\dots 51 \dots 59,~63 & $\pi/4$  & $3\pi/4$ &$\ket{g}$\\
43, 51, 59 & $\pi/8$      & $3\pi/8$ &$\ket{g}$\\
51 & - & - & \\
\hline
\end{tabular}
\qquad
\begin{tabular}{ ||@{\quad}c@{\quad}|@{\quad} c@{\quad} | @{\quad}c@{\quad} |@{\quad} c @{\quad}|| }
\hline
Distilled Fock Nos. & $\phi$ & $\theta$ & $\mathcal{M}$\\
\hline \hline
85,~86\dots100\dots 114,~115 & $\pi$ & 0& $\ket{g}$   \\ 
86,~88\dots100\dots 112,~114 &  $\pi/2$     & 0 &$\ket{g}$  \\  
88,~92\dots100\dots 108,~118 & $\pi/4$      & 0 &$\ket{s}$\\
92, 100, 108 & $\pi/8$      & $\pi/2$ &$\ket{g}$\\
100 & -  & -  & -\\
\hline
\end{tabular}
\caption{Distillation of the Fock states $\ket{51}$ and $\ket{100}$. First column shows the distilled Fock numbers after each iteration. Second column shows the atom-cavity phase, $\mathcal{D}(\phi)$. Other columns show the unitary operation $U_a[\theta]$ and the measurement ($\mathcal{M}$) on the atom.}\label{tab:dist}
\end{table*}
\twocolumngrid
In the main draft, we demonstrated distillation utilizing coherent states. It was evident that the number of iterations required depends on the Fock distribution. However, using squeezed coherent states (SCS) with squeezing can narrow the Fock distribution, potentially leading to a reduction in the required number of iterations. Here, we provide two examples using SCS and compare them with coherent states. These techniques can be applied for a general SCS. First we define the mean and variance of a quantum light 
\begin{align}
\expval{\hat{a}^\dagger \hat{a}} \equiv  \expval{\hat{n}}, &&
\expval{(\Delta\hat{n})^2}  \equiv  \expval{\hat{n}^2}-\expval{\hat{n}}^2, \\
&&Q = \dfrac{\expval{(\Delta\hat{n})^2} - \expval{\hat{n}} }{\expval{\hat{n}}},
\end{align}
When $Q<0$ ($Q>0$), the light mode is said to obey sub (super) Poissonian statistics.
The coherent state is written as  
\begin{align}\label{eq:chs}
\ket{\alpha} = e^{-\dfrac{\abs{\alpha}^2}{2}} \sum_{k=0}^\infty \dfrac{\alpha^k}{\sqrt{k!}}  \ket{k}, &&
 P_k = e^{-\expval{\hat{n}}} \dfrac{\expval{\hat{n}}^k}{k!} ,
\end{align}
where $P(k)$ is the probability for the $k^{\text{th}}$-Fock state and  $\expval{\hat{n}}= \expval{(\Delta\hat{n})^2}   = \abs{\alpha}^2$.  For higher amplitudes $P(k)$ can be approximated with a Gaussian, further three standards are expected to include all the statistics
\begin{align}\label{aqe:cohs}
\sum_{n=0}^\infty  P_k \approx
\sum_{\expval{\hat{n}}-3\expval{\Delta\hat{n}}}^{\expval{\hat{n}}+3\expval{\Delta\hat{n}}} {P_k} = 0.997.
\end{align}
A general SCS is written as \cite{gerry2005}
\begin{align}
\ket{\alpha,\xi} = \hat{\mathcal{D}}(\alpha') \hat{\mathcal{S}}({\xi}) \ket{0}
\end{align}
where $\xi= r e^{\mi \theta }$, $\alpha' = \alpha e^{\mi \phi} $ and $\phi = \theta/2 $ .
$\hat{\mathcal{D}}(\alpha')$ is the displacement operator and $\hat{\mathcal{S}}(r)$ is the squeezing operator.
The probability distribution, variance and the mean are obtained as \cite{gerry2005}
\begin{align}
 \abs{\braket{n}{\alpha',\xi}}^2 \equiv P^{\text{sq}}_k = &~\dfrac{(\frac{1}{2}\tanh r)^k}{k! \cosh r} ~e^{-\alpha^2 (1+\tanh r)} \label{aeq:pp}\\ 
&\qquad \abs{H_k\qty[\dfrac{{\alpha}e^{r}}{\sqrt{\sinh{2r}}}]}^2, \nonumber \\ 
 \expval{(\Delta\hat{n})^2}  = &~ \alpha^2 e^{-2r} + 2 \sinh^2 r \cosh^2 r,\label{aeq:var}\\
 \expval{\hat{n}}=& ~\alpha^2+\sinh^2 r,
\end{align}
where $H_k$ represents the $k^{\text{th}}$ order Hermite polynomial.
with large coherent part ($\alpha \gg \sinh r$) and sub-Poissonian statistics $\expval{(\Delta\hat{n})^2} < \expval{\hat{n}}$, a SCS can give raise to the non-classical effect of number squeezing (see Fig.~\ref{fig:sts})

Similar to the coherent state (\fleq{aqe:cohs}) the SCS  $\ket{10,0.75}$ and  $\ket{\sqrt{51},0.65}$ are numerically verified to saitisfy 
\begin{align}\label{aqe:scoh}
\sum_{n=0}^\infty  P_k^{\text{sq}}  \approx
\sum_{\expval{\hat{n}}-3\expval{\Delta\hat{n}}}^{\expval{\hat{n}}+3\expval{\Delta\hat{n}}}  P_k^{\text{sq}} = 0.9995.
\end{align}

Using \fleq{eq:for} and \fleq{aeq:var}, we can determine that four iterations are required to distill Fock states $\qty{\ket{51}, \ket{100}}$ form $\qty{\ket{10,0.75}, \ket{\sqrt{51},0.65}}$, while using coherent states $\qty{ \ket{10}, \ket{\sqrt{51}} }$ requires five iterations. We performed the distillation protocol to verify this, see  Table.~\ref{tab:dist}. By optimizing over the squeezing parameter, we can reduce the number of iterations by one. However, increasing the squeezing beyond a certain level may cause oscillations in the photon number distribution and require more iterations \cite{gerry2005}, as shown in Fig.~\ref{fig3}.


\bibliography{refs.bib}
\bibliographystyle{apsrev4-2}

\end{document}